\documentclass[epj]{svjour}
\usepackage{graphics}
\usepackage[english]{babel}
\usepackage{psfrag}
\usepackage{slashed}

\begin{document}

\title{Quantitative comparison of filtering methods in lattice QCD}

\author{Falk Bruckmann\inst{1},  Christof Gattringer\inst{2}, 
Ernst-Michael Ilgenfritz\inst{3}, Michael M\"uller-Preussker\inst{3},  
Andreas Sch\"afer\inst{1},
Stefan Solbrig\inst{1} }

\institute{Institut f\"ur Theoretische Physik, Universit\"at Regensburg, D-93040 Regensburg, Germany \and 
Institut f\"ur Physik, FB Theoretische Physik, Universit\"at Graz, A-8010 Graz, Austria \and 
Humboldt-Universit\"at zu Berlin, Institut f\"ur Physik, Newtonstr.\ 15, D-12489 Berlin, Germany}

\date{Received: date / Revised version: date}

\abstract{
We systematically compare filtering methods used to extract 
topological excitations (like instantons, calorons, monopoles and vortices) 
from lattice gauge configurations, 
namely APE-smearing and spectral decompositions based on lattice 
Dirac and Laplace operators.
Each of these techniques introduces ambiguities, 
which can invalidate the interpretation of the results.
We show, however, that
all these methods, when handled with care,
reveal very similar topological structures.
Hence, these common structures are free of 
ambiguities
and faithfully
represent infrared degrees of freedom in the QCD vacuum. 
As an application we discuss an interesting power-law for the clusters 
of filtered topological charge.}


\authorrunning{F.~Bruckmann {\it et al.}}
\titlerunning{Comparison of filtering methods}
\maketitle

\section{Introduction}

Ever since the advent of quantum chromodynamics (QCD)
its infrared properties have been of primary interest.
One of the most important phenomena in this regime is the confinement of quarks.
As a typical nonperturbative effect it still calls for a derivation from first principles,
even for pure gauge theory 
(where it reveals itself as, e.g., an area law for the Wilson loop).

Remarkably, most of the most popular nonperturbative ap\-proach\-es to the QCD vacuum
have been around for quite some time now. 
Topological excitations like instantons, calorons, magnetic monopoles and vortices 
have been used for semiclassical and condensed matter-inspired models 
since the 70's \cite{bruckmann:00c}.
In spite of the successes of these models, 
the question of their relevance for confinement 
is the subject of debate.

Lattice gauge theory (LGT) is of roughly the same age 
and till today the only nonperturbative and systematically improvable regulator of QCD.
Thus it has the potential to lend support to and to 
decide between the various models.

However, the QCD vacuum on the lattice as seen by naive gluonic observables
has been found 
to be dominated by ultraviolet (size $O(a)$) fluctuations and therefore
it is difficult to make contact to continuum models.
To deal with this problem,
various smoothing procedures, filtering out the UV `noise', 
have been developed and applied to lattice configurations.
It has been objected that physical properties could be lost 
and unphysical artefacts may be generated by these filtering methods.    
Both effects would strongly spoil the conclusions,
for instance w.r.t.\ the extracted density of the building blocks,
drawn from any such study.

In this work we systematically investigate the most common 
and a priori quite different procedures to filter lattice configurations,
namely smearing and the modern methods based on the eigenmodes 
of lattice Dirac and Laplace operators.

We find a surprisingly strong agreement of 
the topological content of configurations in thermal equilibrium
seen through the different methods.
We show how the parameters of the latter can be adapted systematically. 
This know\-ledge forms a very important prerequisite 
to uniquely identify the structure of the QCD vacuum.

\begin{figure*}[t]
\begin{center}
\resizebox{0.25\textwidth}{!}{\includegraphics{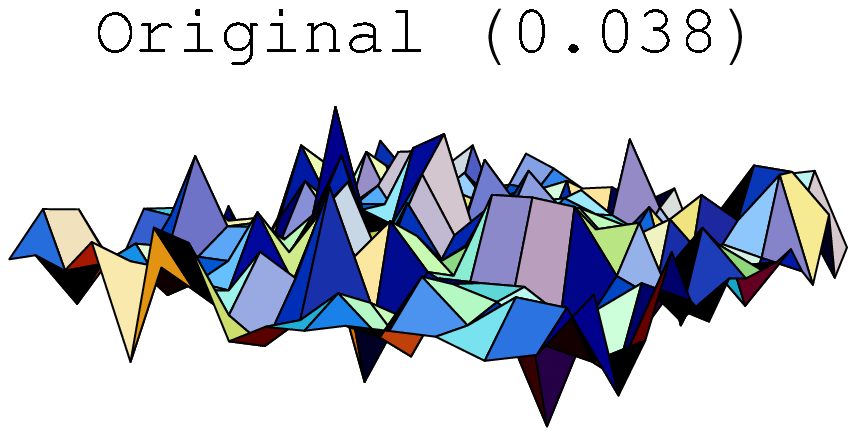}}\\
\vspace*{ 1 cm} 
\resizebox{0.75\textwidth}{!}{\includegraphics{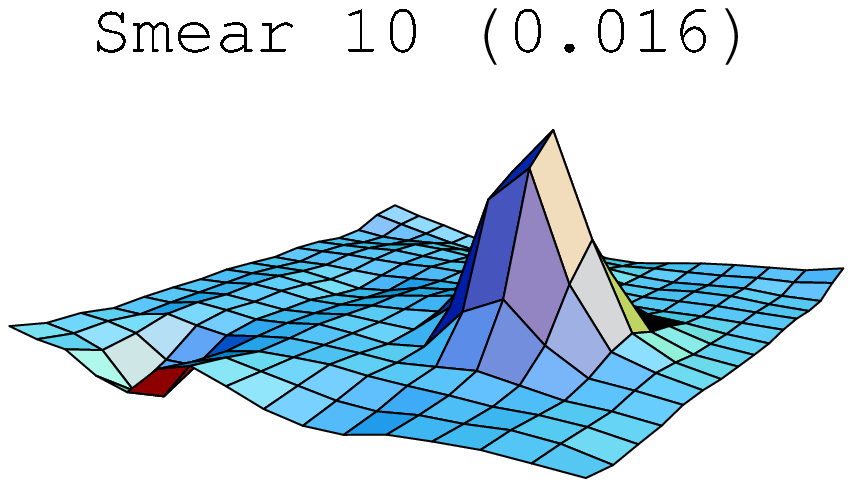}
\includegraphics{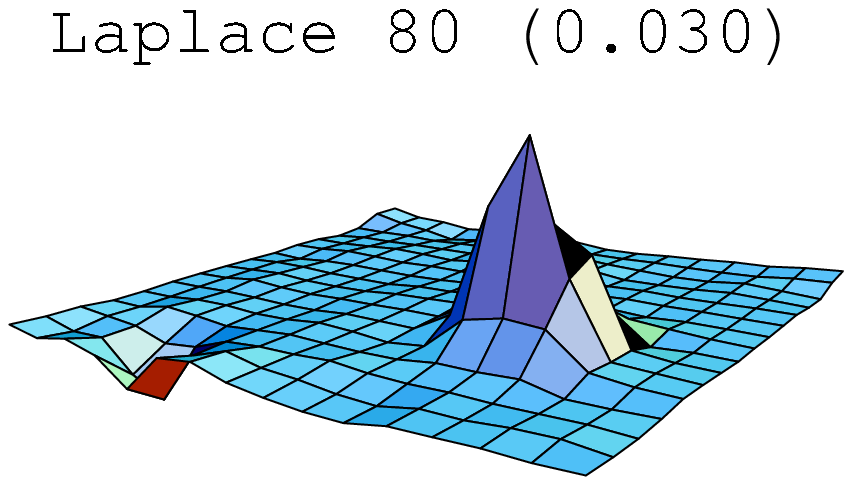}
\includegraphics{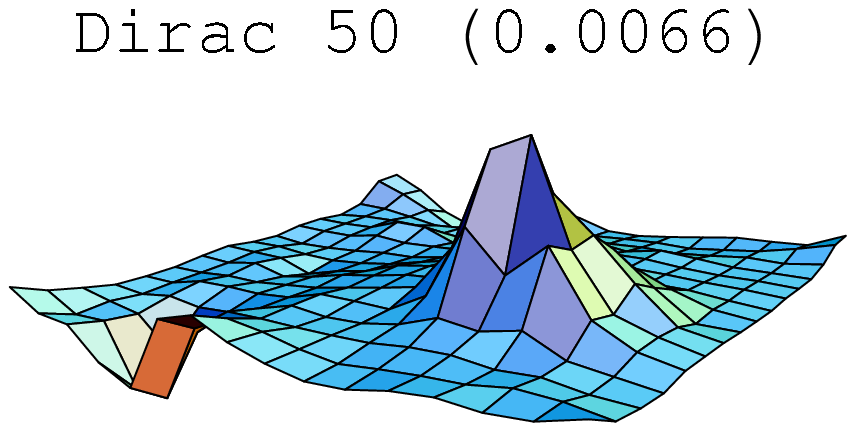}}\\
\vspace*{ 1 cm} 
\resizebox{0.75\textwidth}{!}{
\includegraphics{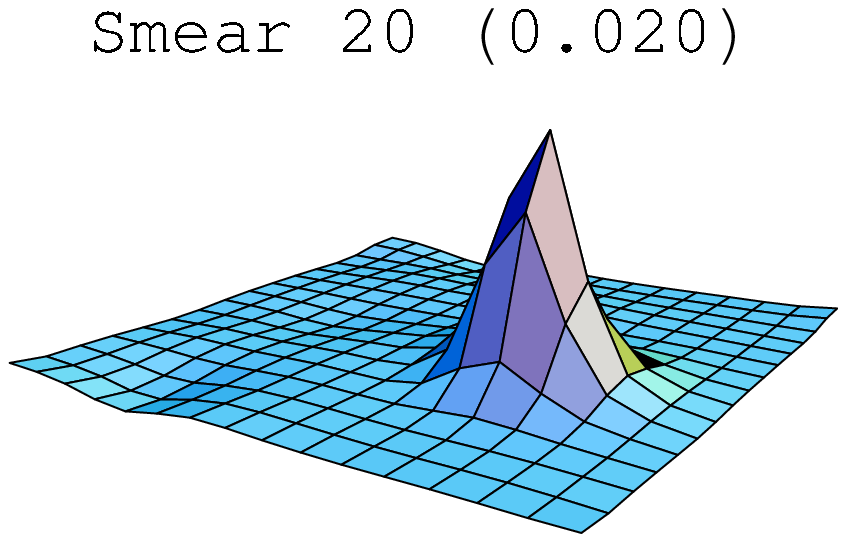}
\includegraphics{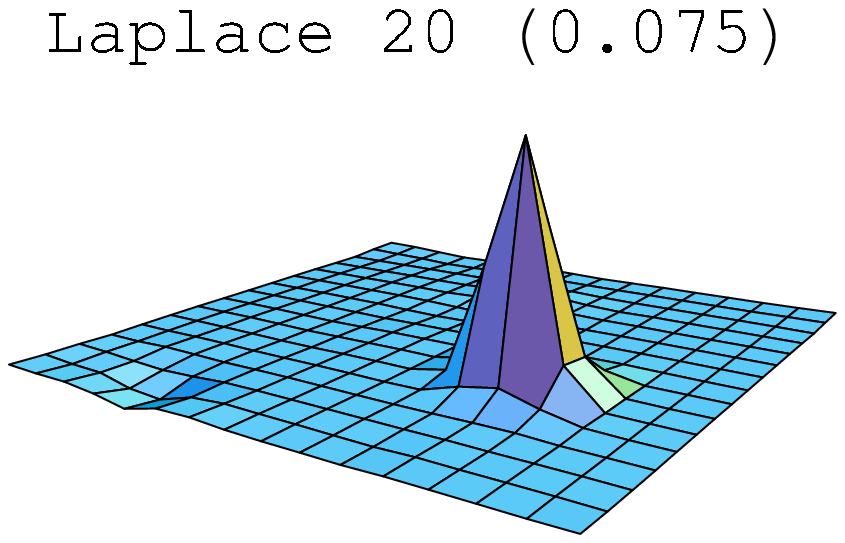}
\includegraphics{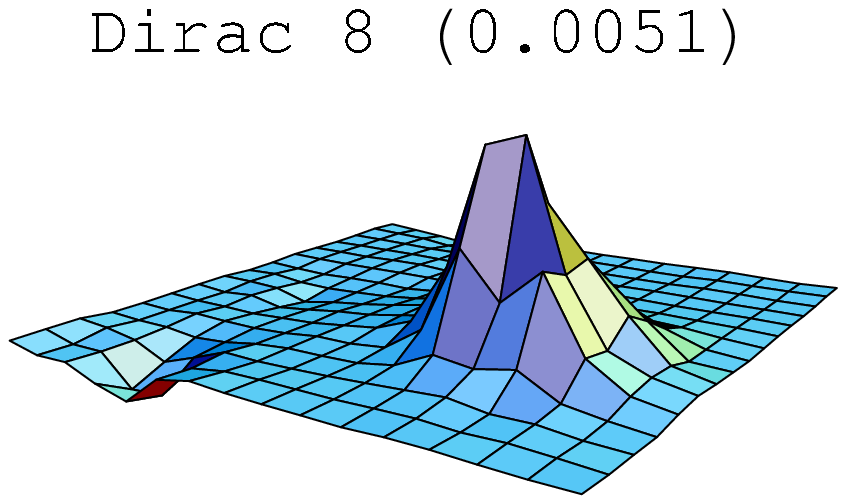}}\\
\vspace*{ 1 cm} 
\resizebox{0.25\textwidth}{!}{\includegraphics{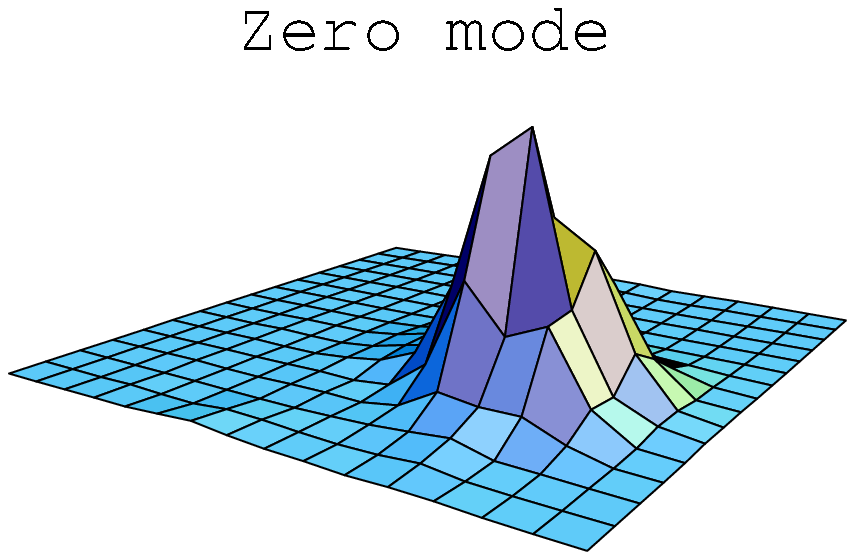}}
\end{center}
\caption{Effect of different filtering methods on the topological density 
for a particular $Q=1$ configuration in a fixed lattice plane. 
On top  we show the original topological density. In the second and third row one sees 
the effects of smearing (first column) and
Laplace filtering (second  column) 
as well as  the topological density in terms of Dirac eigenmodes (last column).
At the bottom the profile of the chiral zero mode is shown for the combined 
filtering methods.
(The numbers in brackets give the heights of the corresponding maxima.} 
\label{fig_showcase_plane}
\end{figure*}

\section{Filtering methods}

{\em Smearing} and the related method of cooling have often been used  
to improve the signal in observables
or to obtain smoother link variables as input for lattice operators.
For definiteness here we will use a 4D version of APE-smearing, 
that has been argued to be equivalent to  
RG cycling \cite{degrand:98b}.
It is an iterative procedure, where links are replaced by a 
weighted average of the links and the staples 
$U_\mu^\nu(x)=U_\nu(x)U_\mu(x+\hat{\nu})U_\nu^\dagger(x+\hat{\mu})$ surrounding it:
\begin{equation}
U_\mu(x)\to\mathcal{P}\big[\alpha U_\mu(x)+\gamma\sum_{\nu\neq \mu} U_\mu^\nu(x)\big]\,.
\end{equation}
Here $\mathcal{P}$ denotes the projection onto the gauge group 
(for $SU(2)$ just a rescaling of the matrix by a scalar).
We choose $\alpha=0.55$ and $\gamma=0.075$, following \cite{degrand:98b}. 
Cooling is obtained by ignoring the old link ($\alpha=0$)
and is known to drive the configurations towards classical solutions.
We update one link at a time, but as the weight of
the central link is quite high, 
our technique qualifies as smearing
rather than cooling.

A more recent idea for filtering is to use eigenmodes of lattice Dirac operators. 
Generally speaking, their function as filters is based on the argument that
low-lying eigenmodes tend to be smooth,
see Fig.\ \ref{fig_showcase_plane} for an example.
The positions revealed by the lowest-lying modes are expected to be correlated 
with the location of the relevant gluonic IR excitations,
in particular of topological objects
\cite{gattringer:06a}. 
Whether peaks show up at the same locations when gluonic filtering methods are applied,
is an important consistency check for the latter.

The {\em Dirac filtering} method relies on the representation of gluonic observables 
through eigenmode expansions of lattice Dirac operators (see also
\cite{gattringer:02c}).
In the following we will 
investigate
the topological charge density \cite{niedermayer:98}
in terms of eigenmodes of a Ginsparg-Wilson type Dirac operator $D$:
\begin{equation}
q(x)={\rm tr}\, \gamma_5(\frac{1}{2}D_{x,x}-1)
=\sum_{n=1}^N (\frac{\lambda_n}{2}-1)\psi^\dagger_n(x)\gamma_5\psi_n(x)
\label{eqn_q_ferm}
\end{equation}
which is exact for $N={\rm Vol}\cdot 4N_c$ and can be truncated for filtering purposes at low $N$
\cite{horvath:03a_ilgenfritz:05b}. 
The total topological charge $Q=\sum_x q(x)$ 
is an integer given by the contributions of chiral zero modes,
whereas the non-zero modes modify the local distribution $q(x)$ only.

The third filtering method we use is the {\em Laplace filter} 
that represents link variables 
through a spectral sum
\cite{bruckmann:05b} 
\begin{equation}
U_\mu(x)=\mathcal{P}\sum_{n=1}^N (16-2\lambda_n)\phi_n(x)\otimes\phi_n^\dagger(x+\hat{\mu})\,.
\end{equation}
Here, $\phi_n$ are the 
eigenmodes of the gauge covariant Laplace operator $-D^2[U]$ 
with eigenvalues $\lambda_n$, of which again only the lowest  
$N$ are taken into account.
In the limit $N={\rm Vol}\cdot N_c$ the original links would be reproduced.
Numerically this filter is cheaper than 
Dirac operators.

The strength of the filters can be controlled.
More iterations of smearing or fewer eigenmodes result in a stronger filtering.
We also remark that smearing is a strictly local procedure, 
while working with eigenmodes introduces an intrinsic nonlocality,
the consequences of which are not fully understood yet.

\section{Results}

We have generated 295 independent quenched $SU(2)$ configurations 
on a $16^4$ lattice with tree-level L\"uscher-Weisz action at $\beta=1.95$ 
(the lattice spacing is \mbox{$a= 0.075(1)\,fm$}).
We use chirally im\-pro\-ved fer\-mions \cite{gattringer:00_00b}, 
an approximate solution of the Gins\-parg-Wil\-son relation, 
which reveal chiral properties well enough
without the big computational demands of, e.g., overlap fermions.

The main physical observable for our comparative study is the topological charge density. 
For smearing and the Laplace method, 
which both provide filtered links,
we use the gluonic definition 
$
q(x)=\sum_{\mu,\nu}{\rm tr}\,
F_{\mu\nu}\tilde{F}_{\mu\nu}/16\pi^2
$
with an improved field strength $F_{\mu\nu}$ 
constructed from $1\!\times\! 1$, $2\!\times\! 2$ and $3\!\times\! 3$ Wilson loops 
\cite{bilson-thompson:02}.
The Dirac eigenmodes yield $q(x)$ directly via Eq. (\ref{eqn_q_ferm}).

In Fig.\ \ref{fig_showcase_plane} we show the result of the three methods
realizing two levels of filtering (mild and strong, see later)
on a thermalized $Q=1$ configuration in a fixed lattice plane. 
It shows an excellent agreement of the `hot spots', i.e.,
lattice locations with large local topological charge, 
visible through the various methods.
For mild filtering also less pronounced structures appear, which agree
between the methods 
(and are washed out by strong filtering).

Moreover, the most pronounced structure
corresponds to a maximum in the profile of the chiral zero mode.
From this consistency 
we can conclude qualitatively that differences between the methods are small.

The configuration used in Fig.\ \ref{fig_showcase_plane} has unit topological charge, 
following from the existence of a single chiral zero mode. 
The following table shows in as far smearing and Laplace filtering recover this fact:

\begin{table}[!h]
\begin{tabular}{lc|cr}
\multicolumn{2}{c}{smearing} &  \multicolumn{2}{c}{Laplace filtering}\\
\hline
sweeps & $Q$ & $Q$ & modes\\
1  & \hspace{0.2cm} 0.948$^*$ \hspace{0.2cm} &  \hspace{0.2cm}1.161$^*$  \hspace{0.2cm}& 160\\
2  & 1.256$^*$ & 0.947$^*$ & 80\\
5  & 1.118$^*$ & 0.896$^*$ & 40\\
10 & 1.004$^*$ & 0.778$^*$ & 20\\
20 & 1.000$^*$ & 0.760$^*$ & 10\\
80 & 1.000$^*$ & -0.138 & 4
\end{tabular}
\end{table}

\noindent The main conclusion here is that 
the three methods 
point to the same total charge $Q=1$. 
Furthermore, $Q$ evaluated after smearing
approaches unity rather quickly and then 
becomes stable.
It also keeps the chiral zero mode (marked with asterisks $^*$ in the table).
The Laplace-filtered configurations, on the other hand, are not arbitrarily smooth, 
such that the gluonic topological charge measured on them is
1 within some margin; 
on the configurations filtered with 10 or more modes 
we find again one chiral zero mode confirming $Q=1$.

The example discussed so far is typical for the whole ensemble 
(including all $Q$-values)
and helps one to visualize the impressive similarity of smearing, 
Laplace filtering and Dirac eigenmodes, which in the following will be quantified.

\begin{figure}[h]
\resizebox{0.5\textwidth}{!}{
\includegraphics{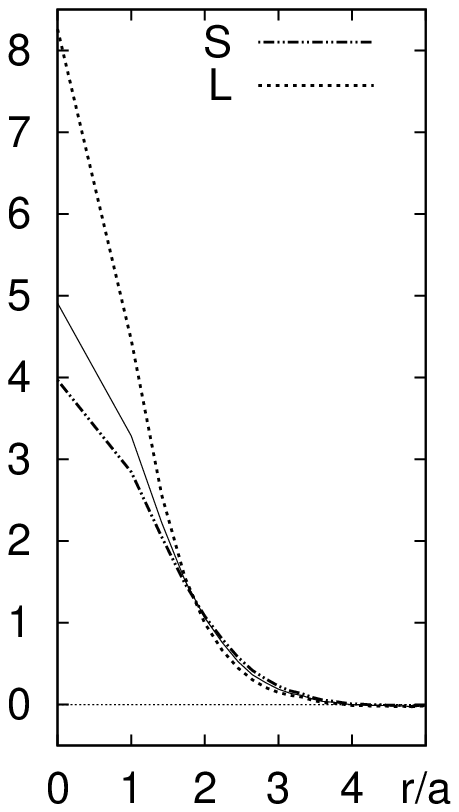}
\includegraphics{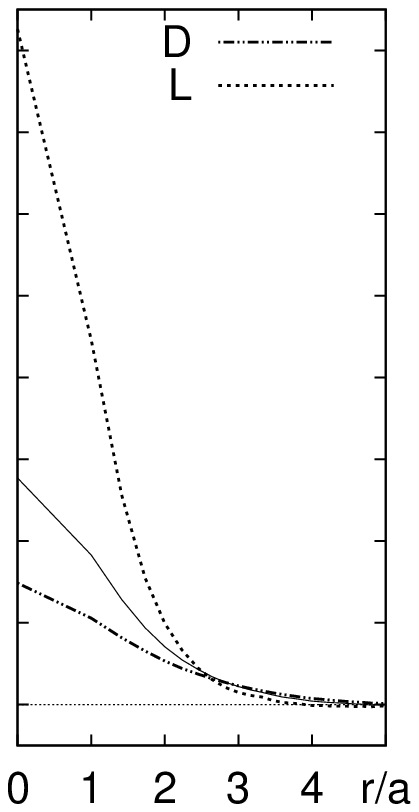}
\includegraphics{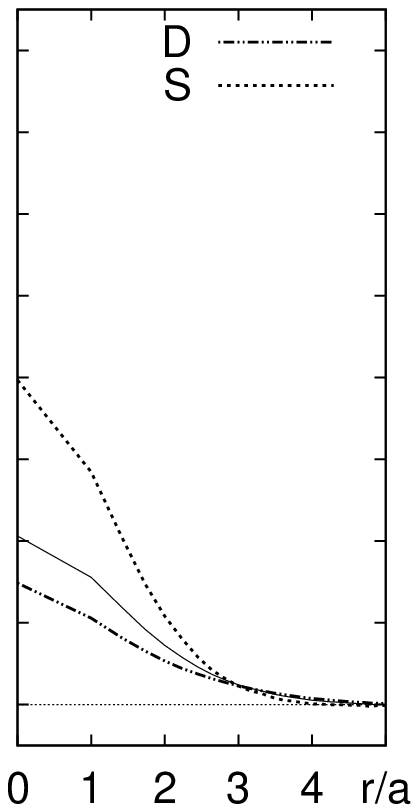}
}
\caption{Auto-correlators $\chi_{AA}(r)$ (broken lines) 
and cross-correlators $\chi_{AB}(r)$ (full line) 
of smearing (S), Laplace (L) and Dirac (D) filtering
for the configuration of Fig.\ \ref{fig_showcase_plane} at mild filtering. 
The unit on the vertical axis is $10^{-7}$.}
\label{fig_chi_r}
\end{figure}

\begin{figure}[!h]
\begin{center}
\resizebox{0.4\textwidth}{!}{\includegraphics{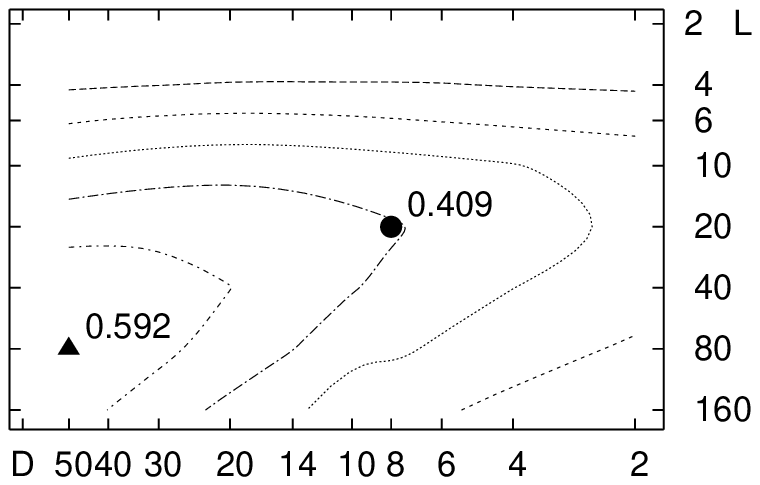}
}\\
\vspace*{0.3 cm}
\resizebox{0.4\textwidth}{!}
{\includegraphics{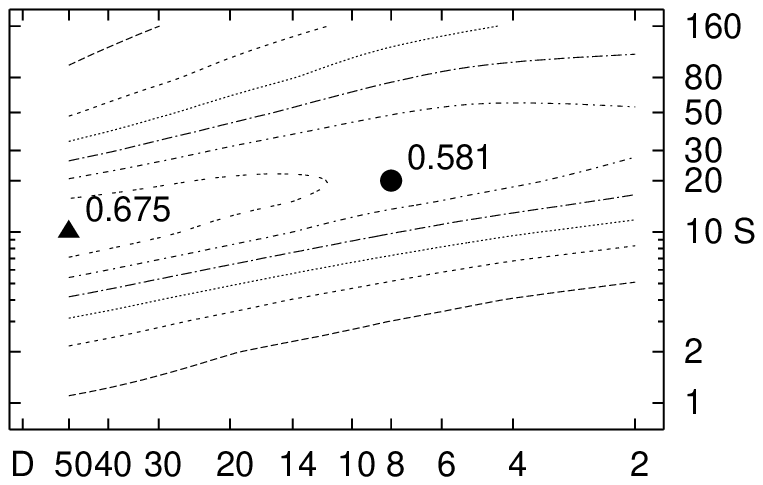}
}\\
\vspace*{0.3 cm}
\resizebox{0.4\textwidth}{!}{
\includegraphics{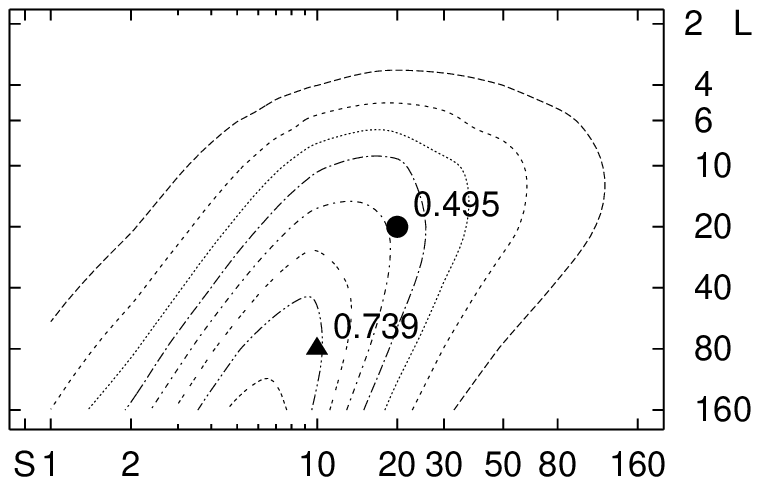}
}
\end{center}
\caption{ 
Level curves $\Xi_{AB}=0.1,0.2,\ldots$ (inwards) for $A$ and $B$ being smearing sweeps (S), 
Laplacian (L) and Dirac (D) eigenmodes, averaged over 10 configurations. 
The parameter sets we use at mild and strong filtering 
are indicated by a full triangle and a circle, respectively.}
\label{fig_chi0_3pairs}
\end{figure}

In particular we try to set the parameters of the filtering methods 
such that the results match as well as possible.
For that purpose 
we consider correlators of the topological charge density 
(with its average $\bar{q}=Q/$Vol subtracted):
\begin{equation}
\chi_{AB}(r)=\frac
{\sum_{x,y}(q_A(x)-\bar{q}_A)(q_B(y)-\bar{q}_B)\delta(|x-y|-r)}
{\sum_{x,y}\delta(|x-y|-r)}
\end{equation}
depending on the four-dimensional distance $r$.
$A$ and $B$ stand for the filtering methods under consideration (including their parameters).

As Fig.\ \ref{fig_chi_r} shows, the auto-correlators $\chi_{AA}(r)$
have a positive profile over a few lattice spacings 
followed by a slightly negative tail, and the cross-correlators $\chi_{AB}(r)$
fall in between them. 
The ratio of the latter to the geometric mean,
$\Xi_{AB}=\chi_{AB}^2(0)/(\chi_{AA}(0)\chi_{BB}(0))$, 
being close to 1 signals local agreement 
between the topological charge landscapes of methods $A$ and $B$.

Fig. \ref{fig_chi0_3pairs} shows $\Xi_{AB}$ for pairwise comparisons 
of smearing, Laplacian and Dirac filtering\footnote{
Investigations that need 50 Dirac modes 
are done on an ensemble of 10 configurations for computational reasons.}.
In all pairs of methods $\Xi_{AB}$ reveals a `ridge' on which the parameters match best. 
From mild filtering (lower left corner\footnote{From the definition it follows
that in the limit of no smearing sweeps vs. 
Vol $\cdot N_C$ Laplacian modes one has $\Xi_{AB}=1$.}) 
to stronger filtering (upper right corner) 
the methods deviate from each other and the height of the ridge decreases. 
Two optimal parameter sets chosen from that figure (and indicated in it) 
are used in due course:

\begin{table}[!h]
\begin{tabular}{rccc}
                           &\hspace*{0.2cm}    smearing \hspace*{0.2cm}      
                           &\hspace*{0.2cm}   Laplace \hspace*{0.2cm}
                           &\hspace*{0.2cm}   Dirac \hspace*{0.2cm}  \\
mild filtering 
\hspace*{0.3cm}            &  10 sweeps & 80 modes & 50 modes\\
strong  filtering
\hspace*{0.3cm}            &  20 sweeps & 20 modes & 8 modes
\end{tabular}
\end{table}

\noindent These matchings have also been confirmed by comparing smeared 
and Laplace-filtered links minimizing
\[
\sum_{x,\mu}{\rm tr}\, (U_\mu^A(x)-U_\mu^B(x))^\dagger(U_\mu^A(x)-U_\mu^B(x))~~~.
\]
For these two optimized parameter sets
we summarize in the following table
some additional measurements, 
which characterize the deviation of the smeared and Laplace-filtered
ensemble from the orginal Monte Carlo one (295 configurations).
The observables are the percentage of configurations
for which the topological charge $Q$ coincides 
with the number of zero modes (called $Q_D$) 
and the decrease of the action $S$: 
\begin{table}[!h]
\begin{tabular}{rcc}
& \hspace*{0.3cm}$|Q-Q_D|\leq 0.5$\hspace*{0.3cm} & 
$ \hspace*{0.3cm}S/S_{\rm orig}$\hspace*{0.3cm} \\
10 sweeps & 89\% & 0.026 \\ 
80 modes  & 76\% & 0.036 \\    
\hline
20 sweeps & 85\% & 0.009 \\ 
20 modes  & 69\% & 0.017 \\ 
\end{tabular}
\end{table}

\noindent In reducing the noise, 
the action has been reduced by approximately two orders of magnitude,
while the string tension is unchanged within errors. 
   
\begin{figure*}[!t]
\begin{center}
\resizebox{1\textwidth}{!}{
\includegraphics{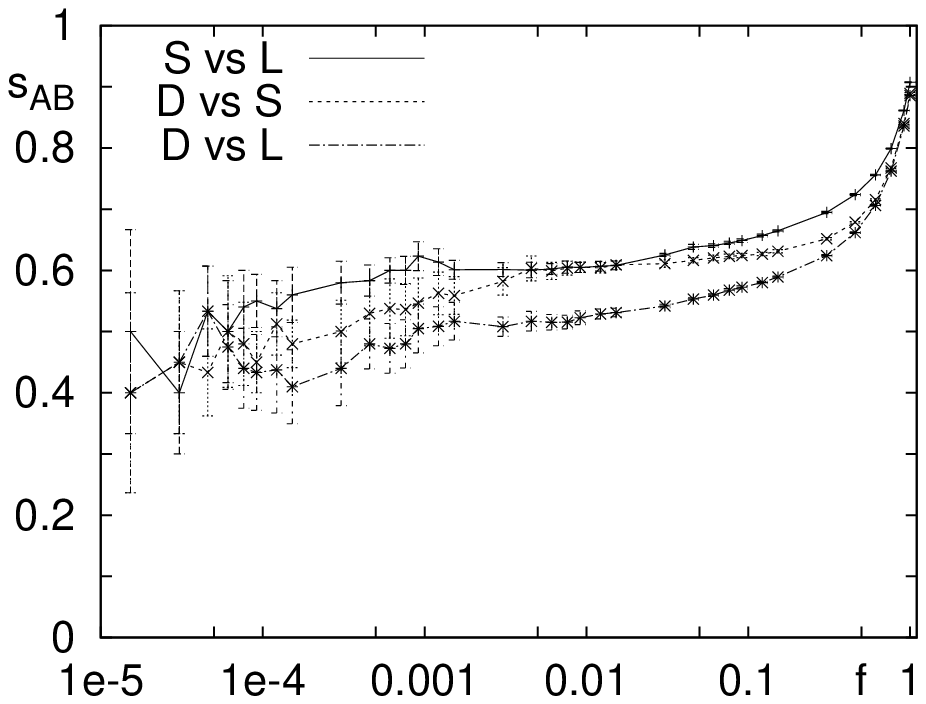}
\includegraphics{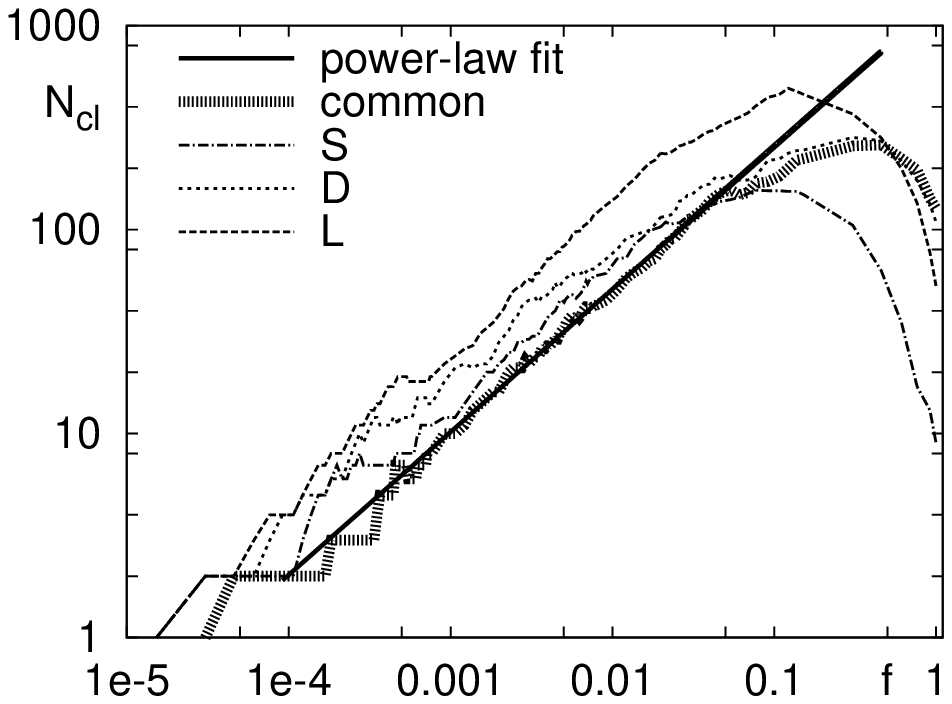}
}
\end{center}
\caption{RPOs averaged over 10 configurations (left) 
and the number of clusters 
for the configuration of Fig.\ \ref{fig_showcase_plane}
(right, 
logarithmically, 
including a power-law fit), both 
as a function of the volume fraction $f$ 
at mild filtering.}
\label{fig_point_overlap}
\end{figure*}

With the parameter sets at hand we further quantify the local agreement of the methods 
by looking at clusters of filtered topological charge.
For each method $A$, cuts in the topological charge density
are adjusted such
that the sets $X_A$ of points above the cut have 
the same volume fraction $f={\rm vol}(X_A)/$Vol.
Then we compare the volume
in the overlap (with same sign of topological charge) vs.\ the union of pairs $X_A$ and $X_B$
to obtain the relative point overlap (RPO) $s_{AB}$:
\begin{equation}
s_{AB}=
    \displaystyle\sum_{
      \stackrel{x\in X_A \cap X_B}
      {q_A(x)q_B(x)>0}}
    1 \:\:/
    \displaystyle\sum_{x\in X_A\cup X_B}\!\!\!\! 1\,.
\end{equation}
In Fig.\ \ref{fig_point_overlap} upper panel the RPOs are plotted 
against the volume fraction $f$. 
The main result to emphasize here is that the pairwise
point overlaps are large, typically 50\% to 60\%, 
and constant over a wide range of $f$, 
which implies that also the shapes of the topological lumps agree.

Based on our analysis we are able to discard 
ambiguities
of the individual methods.
When we analyze only points that are common to all three methods,
the number of clusters is reduced accordingly,
see Fig.\ \ref{fig_point_overlap} lower panel.

Intermediate 
values of $f$
are suited for a cluster analysis 
as they contain enough statistics and avoid cluster mergings. 
Interestingly, in that regime we find
for the common structures a pronounced {\em power-law 
for the number of clusters as a function of} $f$.
In order to better interpret this behavior, we will discuss in the next section
gross features of a class of models 
including the dilute instanton gas 
that predict such a power-law.

\section{A model for the power-law} 

We are now considering models 
of topological lumps in the continuum.
The latter are assumed to be dilute and described by 
an arbitrary shape function $F$ with size parameter $\rho$:
\begin{equation}
\label{eq:qOfSingleObject}
q(r) = F(r/\rho)\;\rho^{-\delta}\,.
\end{equation}
When applying cuts to the topological charge density, it is useful 
to characterize the lumps by their maximum
\begin{equation}
\label{eq:q0}
q_0\equiv q(0)= F(0)  \rho^{-\delta} \,. 
\end{equation}
Let us assume a 
power-law distribution of size parameters $d(\rho)\sim \rho^{\beta}$, 
which
can be translated into a power-law distribution of maxima 
\begin{equation}
\label{eq:rhoPowerLaw}
d(q_0)\sim q_0^{-\alpha}\,,\qquad
\alpha=(\beta+\delta+1)/\delta\,.
\end{equation}
Then the number of clusters visible above a cut $q$ is 
simply
\begin{equation}
\label{eq:Nclust}
N_{\rm clust}(q)=
\int_q^\infty{\rm d} q_0\, d(q_0)\sim q^{1-\alpha}\,.
\end{equation}

The next step is to compute the total volume of these clusters.
We assume that $r$ is a $d$-dimensional radius.
Then for a single cluster 
(with maximum $q_0$) we obtain the following volume above $q$:
\begin{eqnarray}
\label{eq:volAboveThreshold}
V(q_0, q)&=& \int{\rm d}^{d-1}\Omega \int_0^\infty{\rm d} r\; r^{d-1}\;
\theta( q(r)-q) \\
&\sim& \rho^d F^{-1}(q\rho^\delta)\sim q_0^{-d/ \delta}F^{-1}(F(0)q/q_0)\,.\nonumber
\end{eqnarray}  
Hence the total volume or equivalently the total number 
of lattice points with topological charge density  
above the cut $q$ can be calculated to
\begin{equation}
\label{eq:Npoints}
N_{\rm points}(q)
=
\int_q^\infty {\rm d} q_0
d(q_0)
V(q, q_0)\sim
q^{1-\alpha-d/\delta}\,.
\end{equation}
Together with (\ref{eq:rhoPowerLaw},\ref{eq:Nclust}) this predicts the following exponent
\begin{equation}
\label{eq:xiDef}
\xi\equiv \frac
{{\rm d}\log N_{\rm clust}(q)}
{{\rm d}\log N_{\rm points}(q)}
=\frac{1}{1+d/(\beta+1)}\:\:,
\end{equation}
independently of the actual profile $F$.

For the naive dilute instanton ensemble the various parameters are
$d=\delta=4$ and
$\beta=11 N_c/3-5=7/3$, 
yielding $\xi=5/11$ 
(the precise way of cutting off the instanton size distribution at large $\rho$ will not matter
since we are concentrating on hot spots of large topological charge density, i.e.\ small size).

Tests with toy models of lumps show that the detection of the power-law and its coefficient $\xi$
hardly depends on finite volume or finite lattice spacing, as long as the lumps are dilute.
However, a high density of topological objects modifies $\xi$. 
In the extreme case of the topological charge density being pure noise one obtains $\xi=1$, 
as every cluster comes with just one point.

\begin{figure*}[!t]
\vskip 1.5 cm
\resizebox{0.45\textwidth}{!}{
\includegraphics{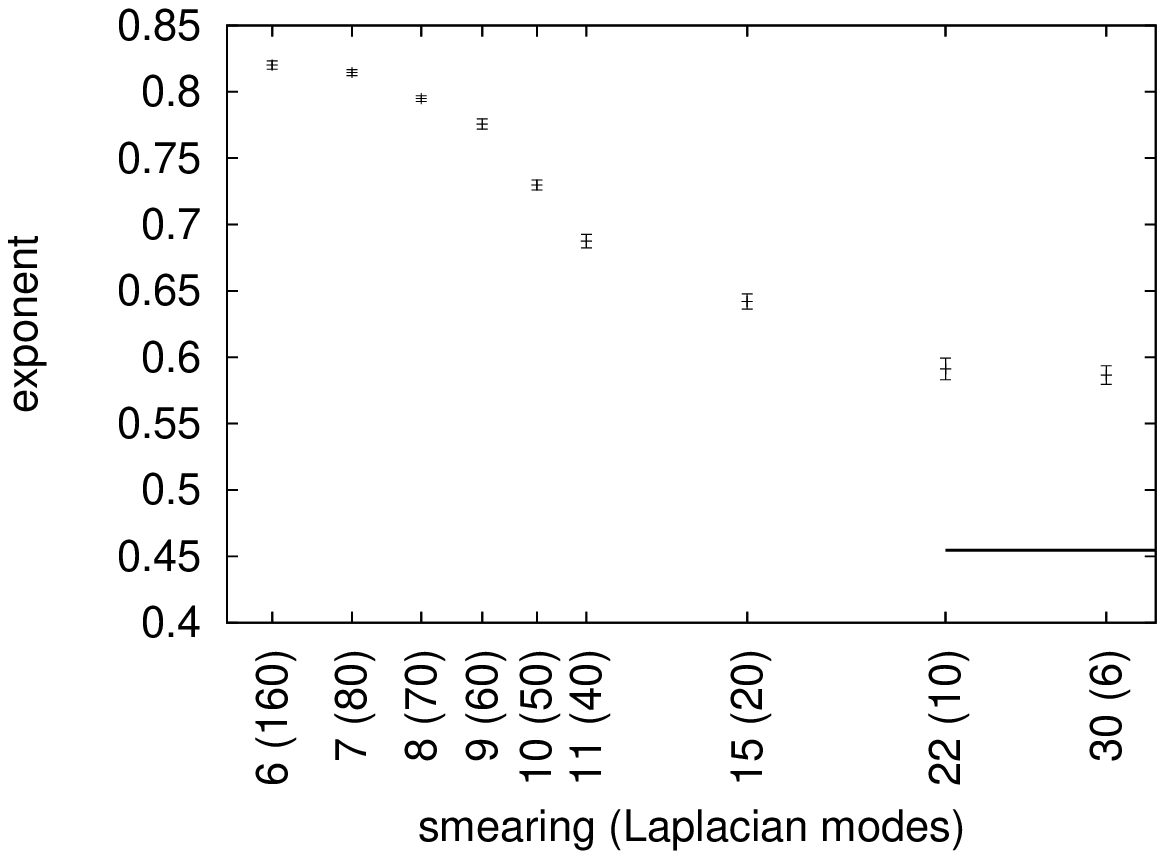}
}
\vskip -5.2 cm
\hskip 8 cm 
\resizebox{0.5\textwidth}{!}{
\includegraphics{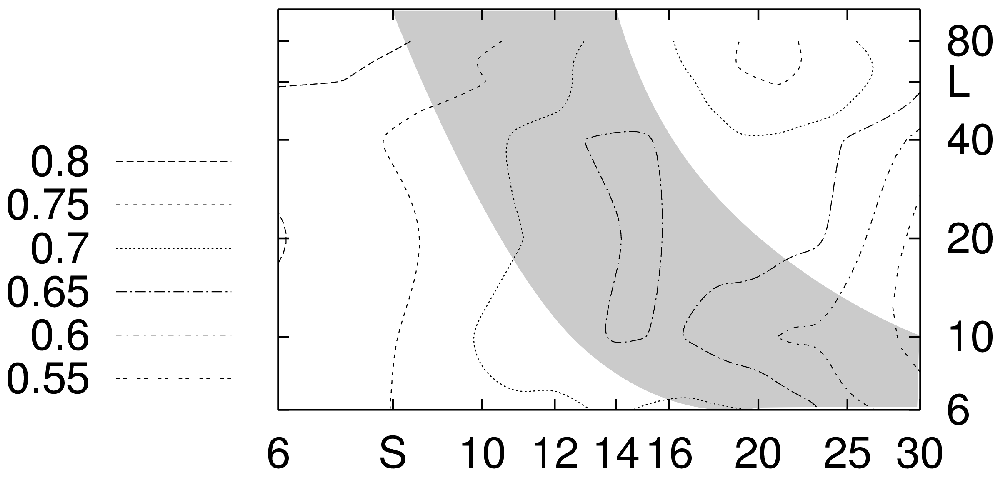}
}
\vskip 1.4 cm
\caption
{The exponent $\xi$ of common topological clusters 
as a function of smearing sweeps and Laplacian modes.
The best matching parameters (cf.\ Fig.\ \ref{fig_chi0_3pairs} right) are used in the left plot 
and marked by a grey band in the right plot.}
\label{fig:powerlawtenconfigs}
\end{figure*}

The measured exponent for the topological charge density 
indeed is close to 1 for very mild or no filtering.
As Fig.\ \ref{fig:powerlawtenconfigs} shows, $\xi$ is lowered 
when removing noise by filtering and tends to a plateau
for moderate filter parameters. 
At strong filtering the cluster statistics becomes unreliable 
since the topological landscape is dominated by very few clusters.

In order to avoid 
ambiguities
of the different methods, 
we have based our analysis on clusters common to smearing and Laplacian modes 
(neglecting the numerically expensive Dirac eigenmodes).
In the upper panel
of Fig.\ \ref{fig:powerlawtenconfigs} we depict $\xi$ along the ridge of optimally matching parameters,
whereas 
in the lower panel of that figure we include other (non-optimal) 
pairs of parameters.

While the statistical errors of the fits are small, 
the systematic errors from choosing the $f$-interval for the power-law are hard to quantify.
Estimating the latter we obtain
the following value of the power-law exponent 
\begin{equation}
\xi=0.59(5)
\end{equation}
for the filtered lumps of topological charge density.

This result certainly excludes the simplest
dilute gas of instantons  
(for which $\xi=0.45$ see above, included in Fig.\ \ref{fig:powerlawtenconfigs} top) 
as primary topological structures. 
This is not surprising, as instanton ensembles are expected to be non-dilute.
For other scenarios the expected behavior depends crucially on the 
specific model assumptions. 
Therefore, we leave it to the proponents of any such model to check
whether it is in agreement with our findings.

\section{Conclusions}

We have presented a systematic comparison of filtering  methods on the lattice, 
namely the gluonic smearing update 
and truncated expansions based on Laplacian and Dirac eigenmodes. 
Our finding, that these filtering methods agree 
surprisingly well, allows us to identify structures of interest 
nearly free of 
ambiguities.

As a general rule, smearing -- which is the cheapest method -- 
represents these objects reasonably well.
In order to control its effects, however, we consider it essential 
to compare several filtering methods. To this end 
we have shown how to match the parameters of the methods optimally.

As a first application we have measured a power-law for the number of filtered clusters 
and presented a class of models in which the corresponding exponent is a function 
of the size distribution coefficient and the dimensionality. 
We have found that the measured exponent cannot be interpreted in terms of a dilute instanton gas.

This work has been supported by DFG 
(Forschergruppe `Gitter-Hadronen-Ph\"anomenologie') and BMBF.


\begin{thebibliography}{10}

\bibitem{bruckmann:00c}
G. 't~Hooft, Saalburg lecture notes (2000)
hep-th/0010225.

\bibitem{degrand:98b}
T.~DeGrand, A.~Hasenfratz, and T.~G.~Kovacs,
Nucl.~Phys. \textbf{B520}, (1998) 301.

\bibitem{gattringer:06a}
C.~Gattringer, E.-M.~Ilgenfritz, and S.~Solbrig,
(2006); hep-lat/0601015.

\bibitem{gattringer:02c}
C.~Gattringer, Phys.~Rev.~Lett. \textbf{88} (2002) 221601.



\bibitem{niedermayer:98}
F.~Niedermayer, Nucl.~Phys.~Proc.~Suppl. \textbf{73} (1999)
105.


\bibitem{horvath:03a_ilgenfritz:05b}
I.~Horvath et~al.,
Phys.~Rev. \textbf{D67} (2003) 011501; 
Phys.~Rev.\textbf{D68} (2003) 114505;\\
E.-M.~Ilgenfritz et~al., Nucl.\ Phys.\ Proc.\ Suppl.
\textbf{153} (2006) 328.

\bibitem{bruckmann:05b}
F.~Bruckmann and E.-M.~Ilgenfritz,
Phys.~Rev. \textbf{D72} (2005) 114502.


\bibitem{gattringer:00_00b}
C.~Gattringer, Phys.~Rev. \textbf{D63} (2001)
114501;\\
C.~Gattringer, I.~Hip, and C.~Lang,
Nucl.~Phys. \textbf{B597} (2001) 451.

\bibitem{bilson-thompson:02}
S.~O.~Bilson-Thompson,
D.~B.~Leinweber,
and A.~G.~Williams, Ann. Phys.
\textbf{304} (2003) 1. 



\end{thebibliography}
\end{document}